\documentclass{mem}
\usepackage{natbib}\usepackage{txfonts}\usepackage{balance}
\usepackage{graphicx}
\usepackage[a4paper]{hyperref}
\usepackage{epstopdf}
\begin{document}
\def\teff{$T\rm_{eff }$}
\def\kms{$\mathrm {km s}^{-1}$}

\title{Modeling the Outskirts of Galaxy Clusters with Cosmological Simulations}

\subtitle{}

\author{Daisuke Nagai}

\institute{
Department of Physics, Yale University, New Haven, CT 06520, U.S.A.\\
\email{daisuke.nagai@yale.edu}
}

\authorrunning{D.Nagai}

\titlerunning{Simulating the Cluster Outskirts}

\abstract{
We present cosmological simulations of galaxy clusters, with focus on the cluster outskirts. We show that large-scale cosmic accretion and mergers produce significant internal gas motions and inhomogeneous gas distribution ("clumpiness") in the intracluster medium (ICM) and introduce biases in measurements of the ICM profiles and the cluster mass. We also show that non-thermal pressure provided by the gas motions is one of the dominant sources of theoretical uncertainties in cosmic microwave background secondary anisotropies. We briefly discuss implications for cluster cosmology and future prospects for understanding the physics of cluster outskirts using computer simulations and multi-wavelength cluster surveys. 
}
\maketitle{}

\section{Introduction}

In recent years, galaxy clusters have emerged as one of the most unique and powerful laboratories for cosmology and astrophysics. Being the largest and most magnificent structures in the Universe, clusters of galaxies serve as excellent tracers of the growth of cosmic structures. Current generation of X-ray cluster surveys have provided independent confirmation of cosmic acceleration and significantly tighten constraints on the nature of dark energy \citep{allen_etal08,vikhlinin_etal09} and alternative theories of gravity \citep[e.g.,][]{schmidt_etal09}. Several ongoing and new X-ray (e.g., {\it eROSITA}) and Sunyaev-Zel'dovich effect (SZE) cluster surveys (e.g., {\it SPT, ACT, Planck}) are underway to improve current cosmological constraints. 

Outskirts of galaxy clusters have special importance for cluster cosmology, because they are believed to be much less susceptible to complicated cluster astrophysics, such as radiative gas cooling, star formation, and energy injection from active galactic nuclei. Dominant physical processes in the outskirts are limited to the gravity-driven collisionless dynamics of dark matter and hydrodynamics of the intracluster medium (ICM). In the hierarchical structure formation model, galaxy clusters grow by accreting clumps and diffuse gas from the surrounding large-scale structure in their outer envelope.  Numerical simulations predict that the large-scale cosmic accretion and mergers give rise to internal gas motions and inhomogeneous gas distribution in the ICM. However, until very recently, observational studies of the ICM have been limited to radii considerably smaller than the virial radius of clusters. 

Recently, {\it Suzaku} X-ray observations have extended X-ray measurements of the ICM profile out to and beyond the virial radius for several clusters \citep{bautz_etal09,george_etal09,reiprich_etal09,hoshino_etal10,kawaharada_etal10}. While these measurements are still quite uncertain, initial results suggested that the observed ICM profiles may deviate significantly from the prediction of hydrodynamical cluster simulations \citep[e.g.,][]{george_etal09}. In addition to testing models of structure formation, these new measurements will be important for controlling systematic uncertainties in cluster-based cosmological measurements. 

In this work, I will present theoretical modeling of the outskirts of galaxy clusters based on cosmological simulations, with highlights on implications for the interpretation of forthcoming multi-wavelength observations of galaxy clusters. The simulations we present here are described in \citet{nagai_etal07a} and \citet{nagai_etal07b}, and we refer the readers to these papers for more details. 

%%%%%%%%%%%%%%%%%%%%%%%%%%%%%
\begin{figure}[t]
\begin{center}
\vspace{-4mm}
\resizebox{\hsize}{!}{\includegraphics[clip=true]{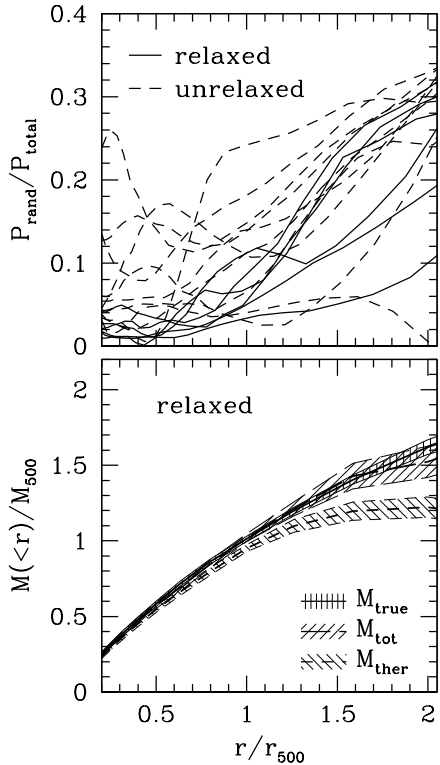}}
\vspace{-8mm}
\caption{\footnotesize
{\it Top panel:} Ratio of pressure from random gas motions to total pressure as a function of radius. Relaxed clusters are represented by solid lines while unrelaxed clusters are represented by dashed lines. {\it Bottom panel:} Averaged mass profiles $M(< r)$ of the relaxed clusters, normalized by $M_{500}$. The solid line shows the actual mass profile from simulation, the long dashed line shows the mass profile from hydrostatic equilibrium including random gas and thermal pressure, and the short dashed line shows the mass profile from hydrostatic equilibrium including thermal pressure only. Hashed region shows the 1-$\sigma$ error of the mean. From \citet{lau_etal09}.}
\vspace{-5mm}
\label{fig:clump_phase}
\end{center}
\end{figure}
%%%%%%%%%%%%%%%%%%%%%%%%%%%%%

\section{Gas Motions in Cluster Outskirts}

Gas motions induced by cosmic accretion and mergers provide non-thermal pressure support in galaxy clusters. The top panel of Fig.~\ref{fig:clump_phase} shows that results of hydrodynamical cluster simulations, illustrating that the non-thermal pressure contributes to 10-20\% at $r=r_{500}$ of the total pressure. The fraction of non-thermal pressure support increases with radius, and it is larger for more dynamically active systems. The bottom panel of Fig.~\ref{fig:clump_phase} shows that the cluster mass profile based on hydrostatic assumption is biased low. We further demonstrate that it is possible correct the bias in the hydrostatic mass, if one could measure gas motions in clusters and hence correct for the bias. Upcoming {\it Astro-H} X-ray satellite mission (scheduled to be launched in 2014) will provide a first direct measurement of gas motions in clusters via doppler broadening of iron lines \citep[][]{inogamov_sunyaev03}.  

%%%%%%%%%%%%%%%%%%%%%%%%%%%%%
\begin{figure}[t]
\begin{center}
\vspace{-4mm}
\resizebox{\hsize}{!}{\includegraphics[clip=true]{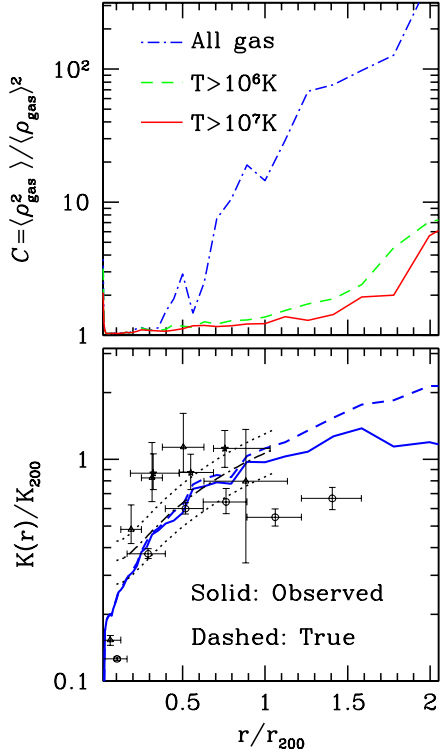}}
\vspace{-8mm}
\caption{\footnotesize
{\it Top panel:} Median clumping factor profiles of gas with different minimum temperature for the simulations with gas cooling and star formation. {\it Bottom panel:} Impact of gas clumping on X-ray measurements of the ICM entropy profile. The dashed line indicates the true entropy profile of the simulated clusters, while the solid line indicates the observed entropy profile, obtained by assuming no clumping ($C=1$). Black points are {\it Suzaku} observations of PKS0745-191 ({\it circles}), A1689 ({\it triangles}), A1413 ({\it stars}), and A1795 ({\it black dot-dashed lines}). From \citet{nagai_etal11}.}\label{fig:clump_phase}
%\vspace{-16mm}
\end{center}
\end{figure}
%%%%%%%%%%%%%%%%%%%%%%%%%%%%%

\section{Gas Clumping in Cluster Outskirts}

Recently, {\it Suzaku} X-ray observations revealed that the observed entropy profile of the ICM is significantly offset from the prediction of hydrodynamical simulations of galaxy clusters. Here, we point out that gas clumping is likely a major source of systematic bias in X-ray measurements of ICM profiles in the envelope of galaxy clusters ($r \gtrsim r_{200}$) \citep{nagai_etal11}. Using hydrodynamical simulations of cluster formation, we show that gas clumping introduces the overestimate of the observed gas density and causes flattening of the entropy profile at large radius.  This is illustrated in Fig.~\ref{fig:clump_phase}. The top panel shows that the clumping factor of the X-ray emitting gas ($T\gtrsim 10^6$~K) is $C \equiv \langle \rho_{\rm gas}^2 \rangle / \langle \rho_{\rm gas} \rangle^2 \sim 1.3$ at $r=r_{200}$, and it increases with radius, reaching $C\sim 5$ at $r= 2 r_{200}$.  In the bottom panel, the solid line indicates the true entropy profile, which is consistent with the self-similar prediction, $K \equiv T/n_e^{2/3}\propto r^{1.1}$ \citep{voit_etal05}. From the definition of entropy, the overestimate of gas density due to clumping causes an underestimate of the observed entropy profile by $C(r)^{1/3}$. Results of our analyses indicate that gas clumping causes the flattening of the observed entropy profiles at $r\gtrsim r_{200}$. While current {\it Suzaku} measurements are still uncertain, our results indicate that gas clumping is important for reducing the tension between recent {\it Suzaku} observations and theoretical prediction of the $\Lambda$CDM model.  

%%%%%%%%%%%%%%%%%%%%%%%%%%%%%
\begin{figure}[t]
\begin{center}
\vspace{-2mm}
\resizebox{\hsize}{!}{\includegraphics[clip=true]{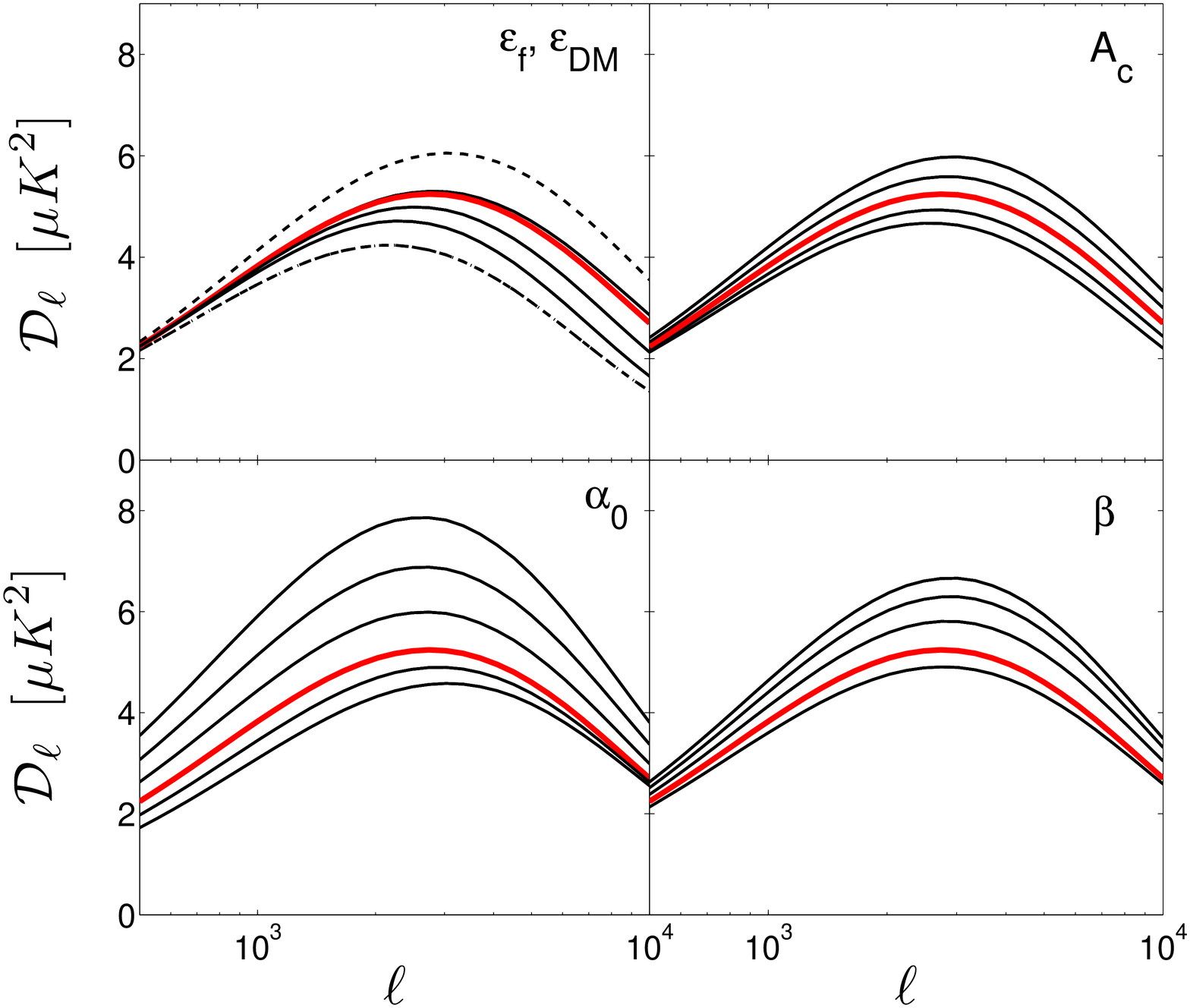}}
\vspace{-2mm}
\caption{\footnotesize
Astrophysical uncertainties in the SZ power spectrum.  Different lines indicate theoretical uncertainties associated with (1) heating of gas by energy injection from stars and AGN (indicated with {\it solid} lines) and merger dynamics (indicated with {\it dashed} lines in {\it top-left} panel), (2) dark matter structures ({\it top-right}), and (3) non-thermal pressure by gas motions ({\it bottom-left}) and its time evolution ({\it bottom-right}). In each case, the thick red line represents our fiducial model. From \citet{shaw_etal10}.}\label{fig:SZpower_model}
%\vspace{-16mm}
\end{center}
\end{figure}
%%%%%%%%%%%%%%%%%%%%%%%%%%%%%

\section{Impact on the SZ power spectrum}

Recent measurements of the SZ power spectrum by {\it SPT} and {\it ACT} telescopes revealed that the SZ power is significantly below the signal predicted by the current cosmic structure formation model \citep{lueker_etal10,shirokoff_etal10,dunkley_etal10}. 

In our recent work, we argued that the current SZ power spectrum template is overestimated by 50-100\%, due to lack of important astrophysical process in theoretical modeling of the SZ power spectrum \citep{shaw_etal10}. Fig.~\ref{fig:SZpower_model} illustrates theoretical uncertainties in the thermal SZ power spectrum. Our model accounts for star formation and energy feedback (from supernovae and active galactic nuclei) as well as radially dependent non-thermal pressure support due to random gas motions, the latter calibrated by recent hydrodynamical simulations. Varying the levels of feedback and non-thermal pressure support can significantly change both the amplitude and shape of the thermal SZ power spectrum. Increasing the feedback suppresses power at small angular scales, shifting the peak of the power spectrum to lower $l$. On the other hand, increasing the non-thermal pressure support significantly reduces power at large angular scales. In general, including non-thermal pressure at the level measured in simulations has a large effect on the power spectrum, reducing the amplitude by $\gtrsim 60$\% at angular scales of a few arcminutes compared to a model without a non-thermal component. Comparing with the recent measurements of the small-scale cosmic microwave background power spectrum, our model reduces the tension between the values of $\sigma_8$ measured from the SZ power spectrum and from cluster abundances.

\section{Future Prospects}

Modern numerical simulations predict that gas motions and clumping are ubiquitous in the outskirts of galaxy clusters. While they are generic predictions of the concordance $\Lambda$CDM model, we have had very little observational handle on these phenomenon until very recently.

New X-ray and SZE observations just coming online have significantly extended measurements of the ICM, out to and beyond the virial radius of clusters. For example, deep X-ray imaging of nearby clusters with current {\it Suzaku} and {\it Swift}-XRT have started to provide accurate measurements of the ICM profiles in the outer envelope of galaxy clusters. Comparison of current X-ray and SZE measurements should soon provide insights into the properties of the ICM in cluster outskirts. {\it Astro-H} X-ray mission promises to provide the first direct measurements of internal gas motions in clusters, and {\it eROSITA} and {\it WXRT} X-ray mission will help increase the number of clusters with deep imaging data extending out to large radii. Theoretical modeling based on detailed numerical simulations is also underway. A plethora of activities (in both theory and observation) will advance our understanding of cluster physics and provide foundation for the use of galaxy clusters as laboratories for cosmology and astrophysics. 

\begin{acknowledgements}
I would like to thank my collaborators Andrey Kravtsov, Erwin Lau, Laurie Shaw for rewarding collaborations which produced results described here. This research was carried out at the Yale University and was supported in part by NSF under grant AST-1009811. 

\end{acknowledgements}

%\bibliographystyle{hapj}
%\bibliography{ms}

\end{document}